# Laser Cooled High-Power Fiber Amplifier

**Galina Nemova**

*Department of Engineering Physics,*
*École Polytechnique de Montréal, P.O. Box 6079, Station Centre-ville, Montréal, Canada*
galina.nemova@polymtl.ca

**Abstract:** A theoretical model for laser cooled continuous-wave fiber amplifier is presented. The amplification process takes place in the $Tm^{3+}$-doped core of the fluoride ZBLAN ($ZrF_4$-$BaF_2$-$LaF_3$-$AlF_3$-NaF) glass fiber. The cooling process takes place in the $Yb^{3+}$:ZBLAN fiber cladding. It is shown that for each value of the pump power and the amplified signal there is a distribution of the concentration of the $Tm^{3+}$ along the length of the fiber amplifier, which provides its athermal operation. The influence of a small deviation in the value of the amplified signal on the temperature of the fiber with the fixed distribution of the $Tm^{3+}$ions in the fiber cladding is investigated.

Heat generated in traditional exothermic solid-state amplifiers or lasers is the source of increased temperature and stress inside the amplifier or laser medium, which cause poor beam quality and limit the average output power. The very high surface-to-volume ratio and guiding in optical fiber amplifiers and lasers provide an excellent solution to compete against other high-power laser technologies based on solid-state bulk lasers, such as for example thin-disk lasers. In spite of tremendous progress in the development of high-power fiber devices, which one could see in recent years, various kinds of limitations are now encountered. Indeed the optical intensities in high-power fiber devices have increased enormously and have almost reached the damage threshold of the material. To solve this problem fiber structures with increased mode areas have been suggested. These novel high power fiber amplifiers are based on the multimode fibers with the large diameter of the fiber core [1] or the amplification process in these fiber amplifiers takes place in a rear-earth doped fiber cladding [2]. But even these schemes are ultimately limited by thermal effects if a high quality beam is required. For fibers with limited chemical stability such for example as fluoride fibers water cooling should be avoided. Lowering of the doping concentration and increase in the length of the fiber makes the cooling easier, but increase nonlinear effects such as stimulated Brillouin scattering (SBS) and stimulated Raman scattering (SRS), which can cause depletion of the amplified signal. In 1999, Steven Bowman suggested a radiation-balanced laser, where lasing is accomplished by offsetting the heat generated from stimulated emission by the anti-Stokes cooling effect [3]. The radiation-balanced laser requires precision control of the pump power at each point along the length laser medium. This is not simple problem in the case of the fiber amplifier or laser.

We present a theoretical model of laser cooled high-power fiber amplifier (Fig.1). The amplification process resulting in heating takes place in the uniformly $Yb^{3+}$-doped core of the ZBLAN fiber. The laser cooling takes place in non-uniformly $Tm^{3+}$-doped fiber cladding. The ZBLAN is chosen as a host, since as low phonon material with phonon energy of ~580 $cm^{-1}$ it is able to support laser cooling of solids. The concentration of the $Tm^{3+}$ ions in the cladding of the fiber is a function of the length of the fiber. It depends on the values of the input pump and signal powers and will be determined later in this paper. The idea of using anti-Stokes fluorescence to cool solid state matter was first proposed by German physicist Peter Pringsheim in 1929 [4]. Experimentally the net radiation cooling by anti-Stokes fluorescence in solid materials was observed for the first time only in 1995 by Epstein's research team in $Yb^{3+}$-doped ZBLANP glass [5].

For simplicity of understanding the scheme of the laser cooled fiber amplifier let us consider first of all the amplification process in a conventional exothermic fiber amplifier with $Yb^{3+}$-doped core and un-doped fiber cladding. The trivalent ytterbium has only two available manifolds, the $^2F_{7/2}$ ground state and $^2F_{5/2}$ excited state. The amplification process in this two-level system in a steady-state condition can be described with well known rate equations described in details in Ref.[6]. In our simulations we consider a multimode fiber structure consisting of a core with a refractive index of $n_{co}$ = 1.52 and radius $R_{co}$ = 25 $\mu m$. It has a numerical aperture NA ≈ 0.21. The radius of the fiber cladding is $R_{cl}$ = 185 $\mu m$. Concentration of the $Yb^{3+}$ ions in the fiber core is $N^{Yb} = 1.42\times10^{18}$ $ions/cm^3$ permitting it to be free from any co-operative effects. The input amplified signal has the wavelength $\lambda_s$ = 1$\mu m$. The pumping wavelength is $\lambda_p$ = 0.98$\mu m$. Both the pump power and the amplified signal are supported by the fundamental fiber core mode. For simplicity's sake let us suppose, the fiber amplifier is mounted in the vacuum chamber and connected with the ambient environment only radiatively following to Stefan-Boltzmann law. It is well known that propagating along the length of the fiber the pump power experiences depletion and the signal power experiences amplification. This process illustrated in Fig.2. The input pump power is very high $P_p$ = 1kW. The input signal power is $P_s$ = 2W. As on can see in Fig.2 the signal power begins to increase intensively around the point located on the distance of ~32 $m$ from the beginning of the fiber. The intense local growth of the amplified signal is accompanied by intense growth of the number of the phonons causing the increased local heating of the fiber amplifier. Fig.3 illustrates the temperature distribution along the length of the fiber amplifier without cooling. The peak of the temperature can cause unwanted effects such as mode mixing or, for very high input pump powers, damage of the fiber.

To avoid this unwanted heating we suggest using the laser cooling of the cladding of the amplifier. In this case the cladding of the fiber has to be doped with $Tm^{3+}$ ions and pumped independently from the core with the high power pump ($P_p^{cool}$) named here as a cooling pump power (Fig.1). It is important to remember that $Tm^{3+}$-doped ZBLANP sample in the first time was laser cooled experimentally to -1.2 K from the room temperature at a pump wavelength $\lambda_p^{cool} = 1.9\ \mu m$ in 2000 [7]. Contrary to $Yb^{3+}$ions, trivalent thulium ions have four manifolds. The ground and first exited manifolds can be used in the laser cooling cycle. It is useful to mention that $^3H_4$ manifold of the $Tm^{3+}$ ion lies 6900 $cm^{-1}$ above the $^3F_4$ manifold and can be populated vie exited-state absorption (ESA) at illumination at wavelengths 1.85 $\mu m < \lambda <$ 1.97 $\mu m$. But as was shown experimentally in the work [7] ESA will not essentially disturb the laser cooling process. The mean fluorescence wavelength of $Tm^{3+}$ ions is $\lambda_f^{cool}$ =1.82 $\mu m$. It determines the range of the wavelengths which provide the laser cooling of the $Tm^{3+}$-doped medium, it is 1.82 $\mu m < \lambda <$ 1.97 $\mu m$. Following to experimental work [7] in our simulations the $Tm^{3+}$-doped fiber cladding will be pumped at the same wavelength $\lambda_p^{cool} = 1.9\ \mu m$. The cooling power generated in the fiber cladding can be calculated using the formula obtained in the paper [8].

$$P_{cool} = \frac{A_{eff}^{cl}\, I_s N^{Tm}\, \sigma_{abs}^{Tm}}{1+\frac{\sigma_{se}^{Tm}}{\sigma_{abs}^{Tm}}+A_{eff}^{cl}\frac{I_s}{P_p^{cool}}}\left(\frac{\lambda_p^{cool}}{\lambda_f^{Tm}}-1\right), \qquad (1)$$

where $I_s = hc\gamma_{rad}^{Tm}/(\lambda_p^{cool}\sigma_{abs}^{Tm})$, $\gamma_{rad}^{Tm}$ = 83 $s^{-1}$ is the radiative relaxation rate of the $Tm^{3+}$ions in ZBLAN at room temperature taken from Ref. [9]. $A_{eff}^{cl}$ is an effective mode area of the cladding mode supporting the cooling pump power in the fiber cladding. $\sigma_{abs}^{Tm}$ and $\sigma_{se}^{Tm}$ are the absorption and the stimulated-emission cross-sections of the $Tm^{3+}$ ions in ZBLAN at the wavelength $\lambda_p^{cool}$ taken from Ref. [9], $N^{Tm}$ is the density of $Tm^{3+}$ ions in the cladding of the fiber, which is a function of the length of the fiber. It will be calculated later in this paper. The cooling pump power $P_p^{cool}$ depleted propagating along the length of the fiber. It can be calculated at any point along the length of the fiber using the equation

$$\frac{P_p^{cool}(0)}{P_p^{cool}(z)} exp\left(\frac{P_p^{cool}(0)-P_p^{cool}(z)}{P_{Sat}^{cool}}\right) - exp(\sigma_{abs}^{Tm} N^{Tm} z) = 0, \quad (2)$$

where $P_{Sat}^{cool} = hc\gamma_{rad}^{Tm} A^{cl}_{eff} /[\lambda_p (\sigma_{abs}^{Tm} + \sigma_{se}^{Tm})]$, $P_p^{cool}(0)$ and $P_p^{cool}(z)$ are the input cooling pump power and the cooling pump power at the point $z$ along the length of the fiber, respectively.

Taking into account Eq. (1) describing the cooling power generated in the fiber cladding and Eq. (2) describing the depletion of the cooling pump power along the length of the fiber, one can easy to calculate the temperature of the fiber ($T_f(z)$) at any point $z$ along the length of the fiber from the equation:

$$2\pi R_{cl} \epsilon \sigma_B \left(T_r^4 - T_f^4(z)\right) = P_{cool}(z) - P_{heat}(z), \quad (3)$$

where, $P_{heat}$ is heating power generated in the fiber core as a result of the amplification process, $T_r$ is the room temperature. The heating power is $P_{heat} = QA_{eff}^{co}$ with the heating power density equal to $Q = h\nu_p W_p - h\nu_s W_s - h\nu_f^{Yb} N_2^{Yb} \gamma_{rad}^{Yb}$. In this equation $\nu_p$ and $\nu_s$ are frequencies of the pump and amplified signal in the core of the fiber, $\nu_f^{Yb}$ is the mean fluorescence frequency of the $Yb^{3+}$ ions in ZBLAN, $N_2^{Yb}$ is the population of $^2F_{5/2}$ excited manifold of the $Yb^{3+}$ ion. $\gamma_{rad}^{Yb}$ = 526 $s^{-1}$ is the radiative relaxation rate of the $Yb^{3+}$ions in ZBLAN at room temperature taken from Ref. [10]. $W_p$ is a pump rate and $W_s$ is a stimulated emission rate. The equations for $W_p$ and $W_s$ are well known and can be found for example in Ref. [6]. $\epsilon = 0.56$ is the hemispherical emissivity of the sample, $\sigma_B$ is the Stefan-Boltzmann constant.

If we want to support the temperature of the fiber constant along all its length and equal for example to the room temperature we need to put $T_f = T_r$ in Eq. (3) and calculate the concentration of the $Tm^{3+}$ ions ($N^{Tm}$) as a function of the length of the fiber. The concentration of the $Tm^{3+}$ ions distributed in such a manner in the fiber cladding along the length of the fiber can provide cooling of the fiber cladding compensating the heat generated in the fiber core during the amplification process. The concentration of the $Tm^{3+}$ ions in the fiber cladding as a function of the fiber length, which permits to support the temperature of the fiber amplifier constant and equal to the room temperature along all length of the fiber during the amplification process is presented in Fig.4 for the input cooling pomp power $P_p^{cool}(0) = 1$ kW.

In all our simulations the input signal power was equal to $P_s = 2$W. Let see what happens with the temperature of the fiber when the input signal power slightly deviates from this value. Of cause the distribution of the $Tm^{3+}$ ions in the fiber cladding illustrated in Fig.4 continues to be invariable in this case. As soon as the input power of the amplified signal changes and becomes equal to $P_s = 1.8$W or $P_s = 2.2$W the distribution of the temperature along the length of the fiber changes too. The temperatures of the fiber in both these cases are presented in Fig.5. As one can see in Fig.5 small changes in the input power of the amplified signal cause small deviations of the temperature of the fiber from the room temperature along its length, but these deviations are not as large as in the case of the un-cooled fiber amplifier illustrated in Fig.3 and can not cause serious problem. If the input power of the amplified signal is $P_s = 1.8$W the differences between the maximum and minimum of the distribution of the temperature along the length of the fiber is $\Delta T \approx 40$K. If $P_s = 2.2$W it is $\Delta T \approx 33$K. This deviation in the temperature of the fiber from the room temperature with the change of the input signal power is tightly connected with the shift of the temperature peak generated by the amplification process along the length of the fiber with the change of the input signal power. This deviation in the temperature can be almost totally compensated slightly changing the length of the fiber. We want to emphasise that our scheme is valid for the amplified signals with the bandwidth $\Delta\nu_s > 8\ GHz$, where it is free from such nonlinear effects as SBS or SRS.

In summary, we have presented a theoretical scheme of a novel laser cooled high power fiber amplifier. For the first time to our knowledge it has been shown that laser cooled cladding with non-uniformly distributed $Tm^{3+}$ ions can compensate (theoretically totally) the heat generated during the amplification process in the uniformly $Yb^{3+}$-doped core of the fiber amplifier for the fixed input pump and signal powers. Small deviation in the power of the amplified signal will cause small deviation in the temperature of the fiber, which will not cause unwanted effects connected with the high local change of the temperature of the fiber. Without loss of generality this laser cooled amplifier can be realised in another low phonon host materials and with another ions, which can support laser cooling.

**References**

[1] D. Taverner, D. J. Richardson, L. Dong, J. E. Caplen, K. Williams, and R. V. Penty, "158-μJ pulses from a single-transverse-mode, large-mode-area erbium-doped fiber amplifier," Opt. Lett. **22**, 378 (1997).

[2] G. Nemova and R. Kashyap, "High-power long period grating assisted EDFA", J. Opt. Soc. Am. B **25**, 1322 (2008).

[3] S. R. Bowman, "Laser without internal heat generation", J. Lightwave Technol. **35**, 115 (1999).

[4] P. Pringsheim, "Zwei bemerkungen über den unterschied von lumineszenzund temperature-strahlung", Z. Phys. **57**, 739-746 (1929).

[5] R. I. Epstein, M. I. Buchwald, B. C. Edwards, T. R. Gosnell and C. E. Mungan, "Observation of laser- induced fluorescent cooling of a solid", Nature (London) **377**, 500-502 (1995).

[6] P. C. Becker, N. A. Olsson, J. B. Simpson, *Erbium-Doped Fiber Amplifiers* (Academic Press, San Diego, 1999).

[7] R.I. Epstein, B.C. Edwards, nad J.E. Anderson, "Observation of anti-Stokes Fluorescence cooling in Thulium-doped glass", Phy. Rev. Lett. **85**, 3600-3603 (2000).

[8] T.R. Gosnell, "Laser cooling of a solid by 65 K starting from room temperature", Opt. Lett. **24**, 1041-1043 (1999).

[9] T. Sakamoto, M. Shimizu, M. Yamada, T. Kanamori, Y. Ohishi, Y. Terunuma, and S. Sudo,"35-dB gain Tm-doped ZBLZN fiber amplifier operating at 1.65 μm", IEEE Photon. Technol. Lett. 8, 349-351 (1996).

[10] J.Y Allain, M. Monerie, and H. Poignant, "Ytterbium-doped fluoride fibre laser operating at 1.02 μm", Electron. Lett. **28**, 988-989 (1992).

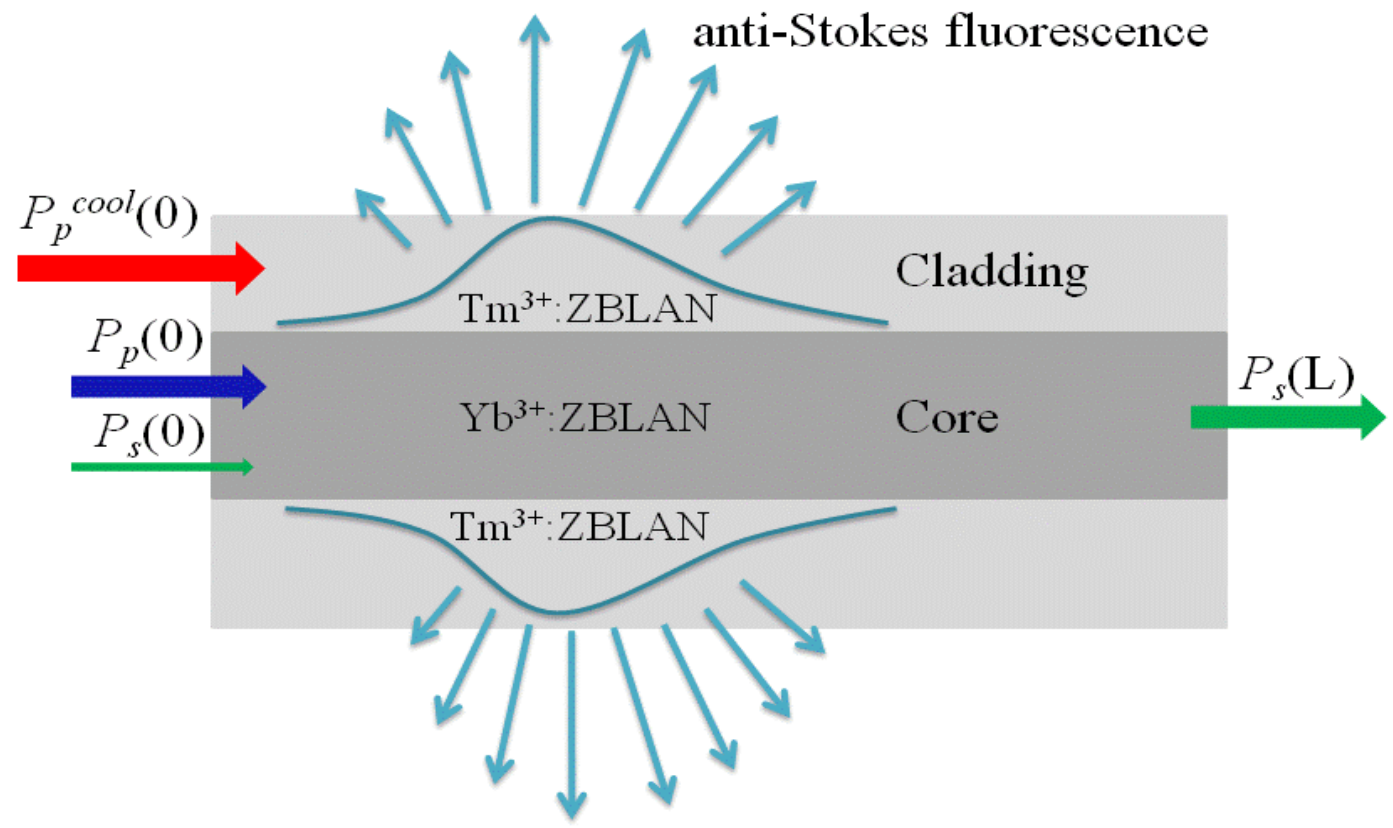

Fig.1. Structure under consideration. The curves in the cladding illustrate the $Tm^{3+}$ distribution.

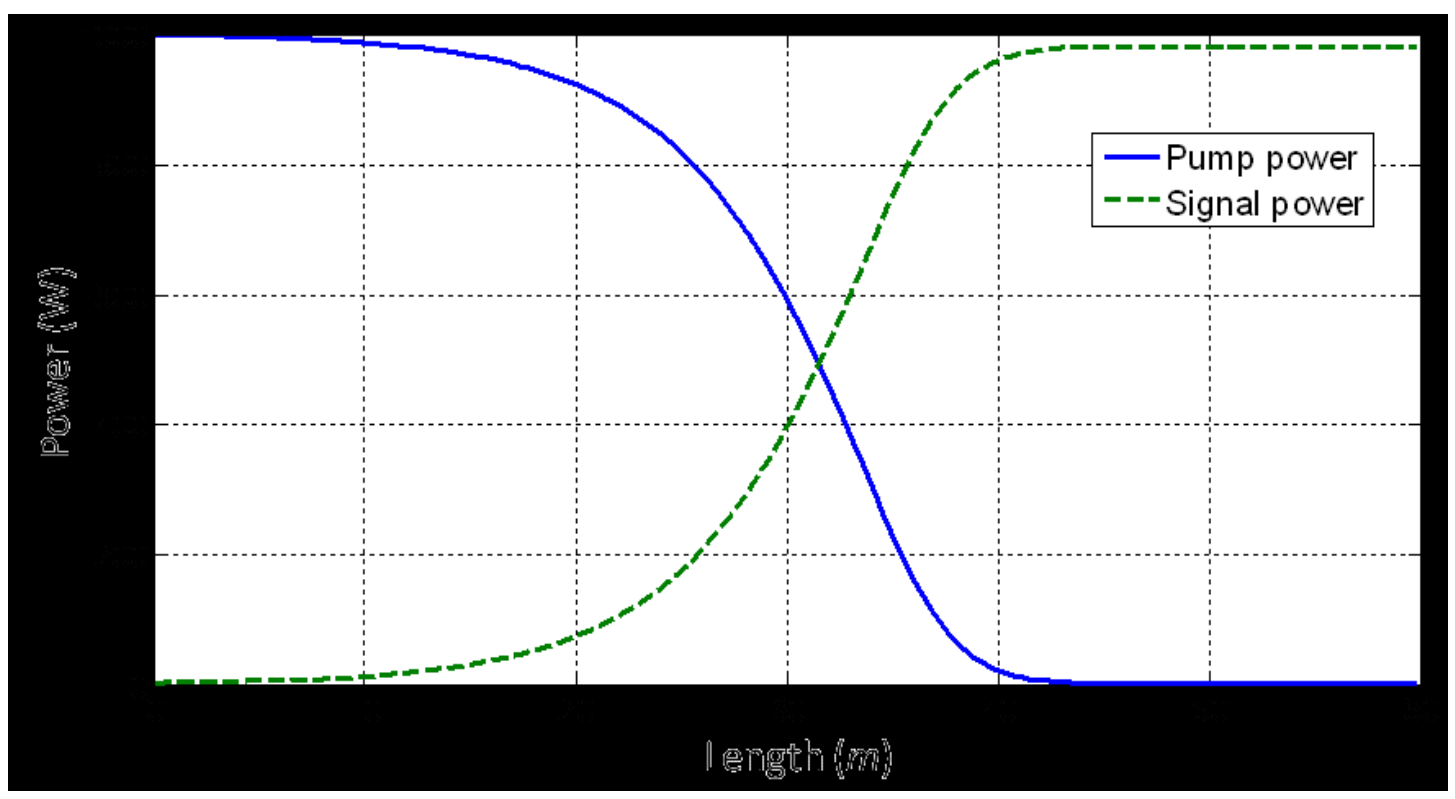

Fig.2. Pump power deletion and signal power amplification along the length of the fiber amplifier.

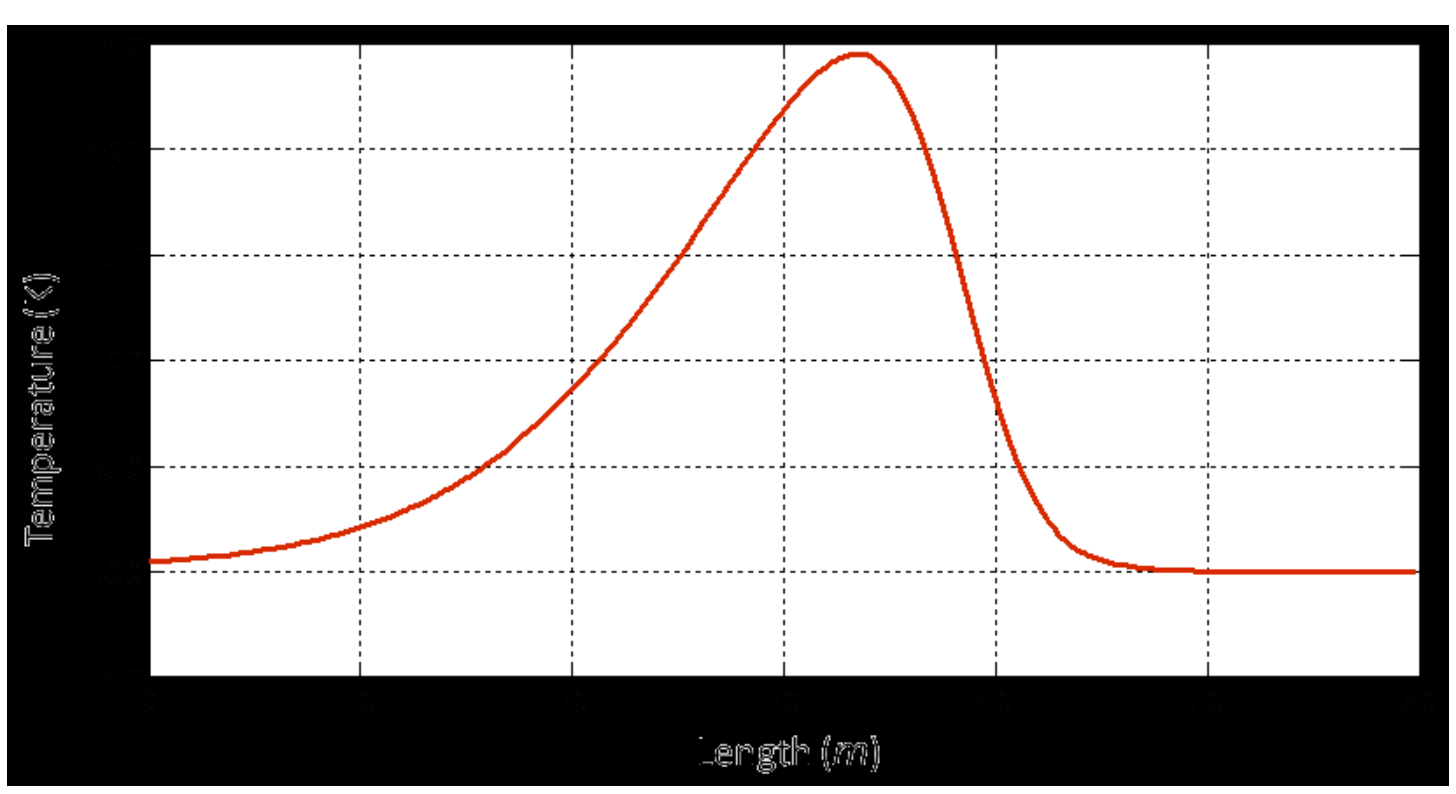

Fig.3. Distribution of the temperature of the fiber along the length of the fiber amplifier, when the laser cooling in the cladding is absent.

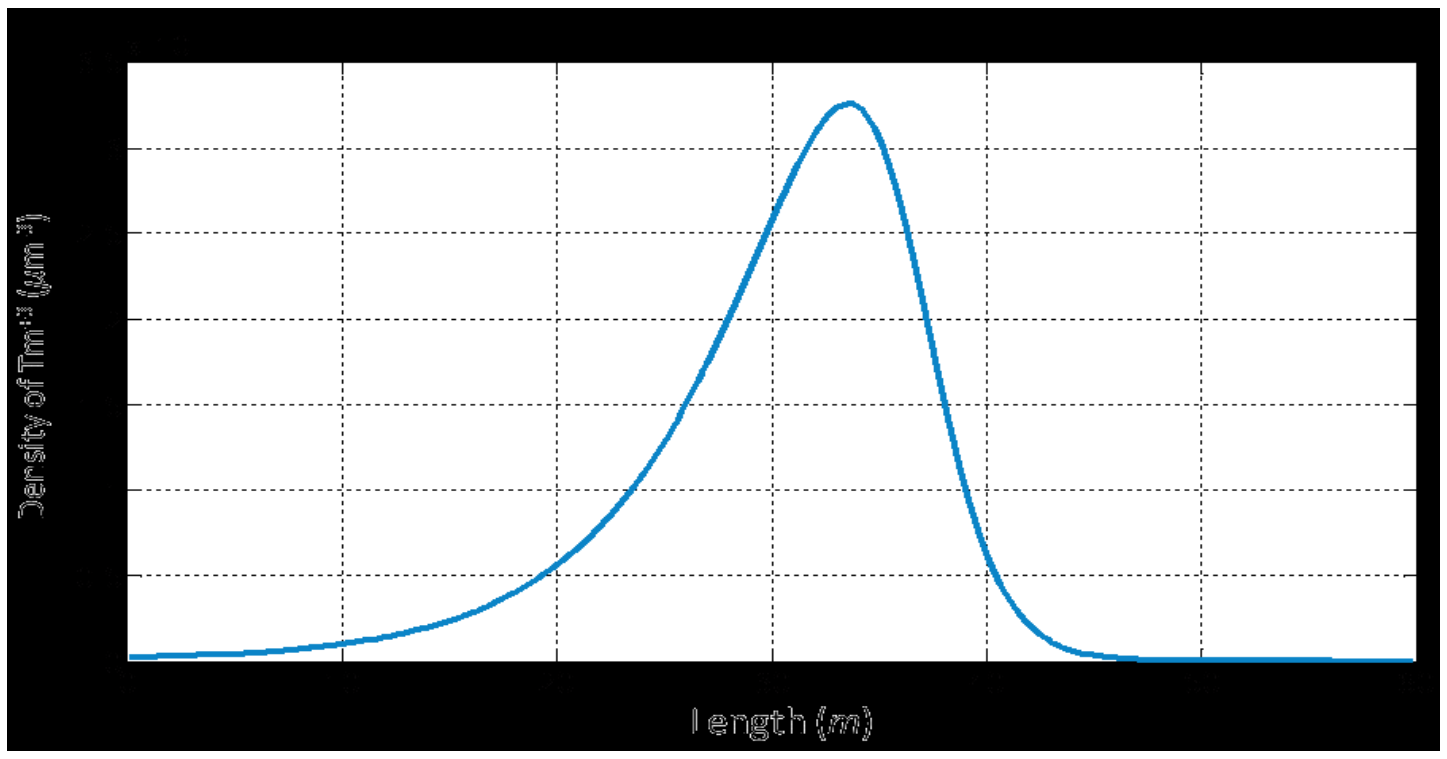

Fig.4. Concentration of the $Tm^{3+}$ ions in the cladding of the fiber as a function of the length of the fiber.

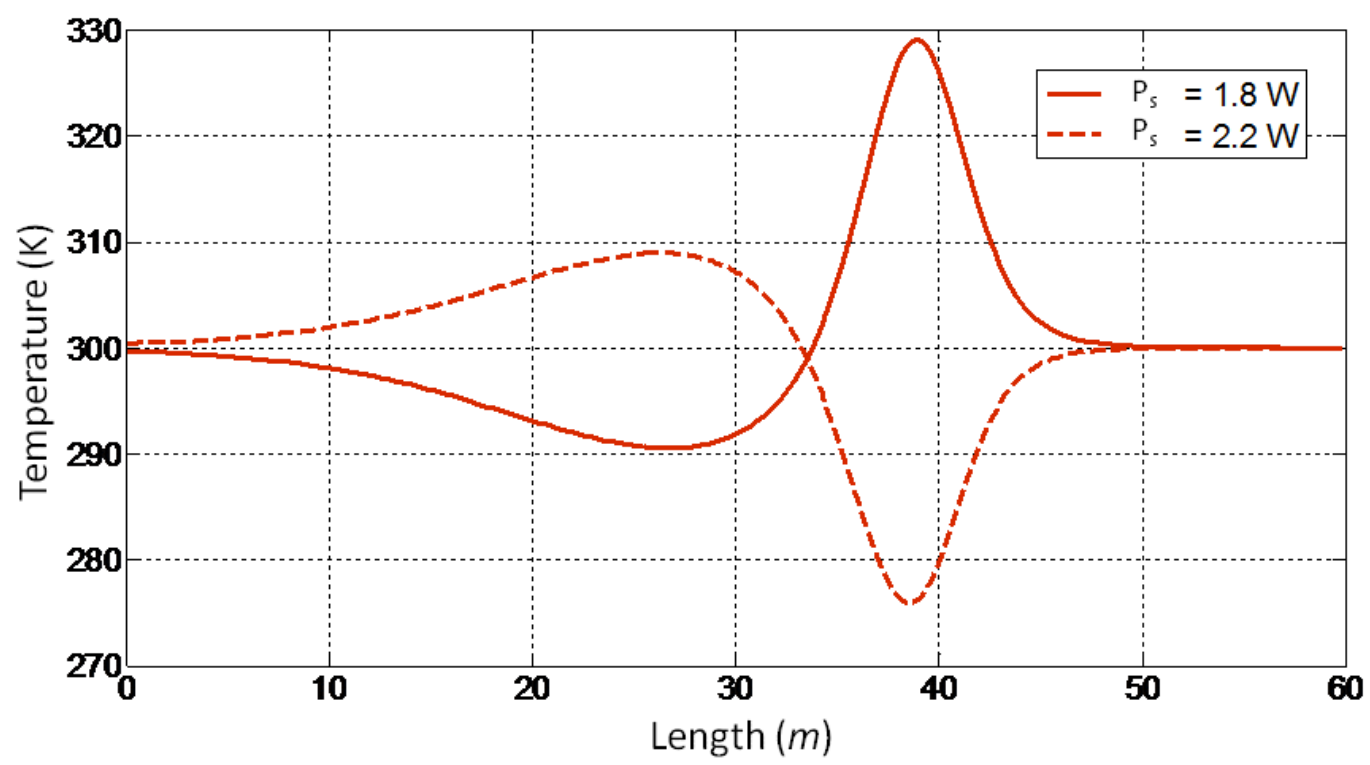

Fig.5. Distribution of the temperature of the fiber along the length of the fiber amplifier for two input powers of the amplified signals ($P_s$). The $Tm^{3+}$-doped cladding pumped with $P_p^{cool}(0) = 1$ kW.